\documentclass[12pt]{iopart}

\usepackage{iopams}  
\usepackage[xdvi]{graphicx}%

\newcommand{\bv}[1]{{\boldsymbol #1}}

\newcommand{\bra}{\left\langle}
\newcommand{\ket}{\right\rangle}

\newcommand{\der}[2]{\frac{d #1}{d  #2}}
\newcommand{\lr}[1]{\left(#1\right)}

\begin{document}
\title[
Non-ergodic transitions in many-body Langevin systems]
{Non-ergodic transitions in many-body Langevin systems: a method of dynamical system reduction}

\author{Mami Iwata}
\ead{iwata@jiro.c.u-tokyo.ac.jp}
\author{Shin-ichi Sasa}
\ead{sasa@jiro.c.u-tokyo.ac.jp}

\address{Department of Pure and Applied Sciences,  
University of Tokyo, Komaba, Tokyo 153-8902, Japan}

\begin{abstract}
We study a non-ergodic transition in a many-body Langevin system.
We first derive an equation for the two-point time correlation function 
of density fluctuations, ignoring the contributions of the third- and 
fourth-order cumulants. For this equation, with the
average density fixed,  we find that there is a critical 
temperature at which the qualitative nature of the trajectories 
around the trivial solution changes. Using a method of dynamical 
system reduction around the critical temperature, we simplify the 
equation for the time correlation function into a two-dimensional 
ordinary differential equation. Analyzing this differential equation, 
we demonstrate that a non-ergodic transition occurs at some temperature 
slightly higher than the critical temperature.
\end{abstract}

\maketitle

\section{Introduction} 

%
%

Glassy systems exhibit distinctive phenomena, such as divergence of the
viscosity \cite{angell}, a history-dependent response \cite{history}, and an extreme slowing down 
of the time correlation of density fluctuations \cite{gotze, ergodic}. These phenomena have been studied extensively through laboratory experiments, numerical experiments, and theoretical analyses of models \cite{kob}.
Although each of these phenomena has gradually come to be understood individually, we are still lacking a unified picture that can account for all of them.
In contrast to the rich variety of physical phenomena displayed by these systems, the types of theoretical methods that have been used in their investigation are few.
Particularly, 
many analyses are based on the mode coupling theory (MCT) \cite{gotze}. 
However, the MCT is not justified completely from a microscopic viewpoint, and also it is not regarded as the first-order description of a systematic formulation.
Therefore, it is important to formulate a new 
method for analyzing the phenomena of glassy systems. 

%
%
Recalling
the historical development of the theory of glassy 
systems, we realize that the MCT has been the most commonly used theory \cite{das} 
because it describes various phenomena on the basis of the singular behavior of the two-point time correlation function of density fluctuations. 
This time correlation function exhibits a two-step 
relaxation as a function of the time difference and a plateau regime appears 
between the two steps in the relaxation. The MCT predicts the transition 
temperature for a fixed density (or the transition density for a fixed 
temperature) at which the plateau regime extends to infinity.
This phenomenon is called a {\it non-ergodic transition}. Although 
the singularity might be an artifact of the approximation employed
in the theory \cite{das}, there are evidences suggesting that the non-ergodic
transition has been observed experimentally and numerically with fairly 
good quantitative correspondence to the theoretical prediction \cite{ergodic}.
Noting such evidence, in the present paper, we theoretically study
the behavior of the time correlation of density fluctuations.

%
%

Let us briefly review the essence of the MCT. 
This theory provides a self-consistent integral equation 
of the time correlation function and the response function, ignoring
 vertex corrections \cite{gotze}.
Then, by analyzing this equation numerically, one can find 
a non-ergodic transition. Here, it should be noted that the memory contribution to the equation for the time 
correlation function is expressed in terms of the time 
correlation function. It is believed that such non-linear 
memory plays a key role in the non-ergodic transition. 

%
%

Now, it is interesting that behavior similar to that exhibited in 
the non-ergodic transition seen in glassy systems 
is easily obtained in ordinary differential equations possessing no 
nonlinear memory. As an example, let us consider 
Newton's equation of motion for a point particle moving in a one-dimensional 
space. Let $x(t)$ be its position and $V(x)$ be a potential function 
given by $V(x)=-x^2\left[(x - x_*)^2+\varepsilon\right]$, with $\varepsilon \ge 0$. We consider the particle motion under the condition that $x \to 0$ as $t \to \infty$. With this condition, it is easily found that when  $\varepsilon$ is small,
the particle starting from $x(0)$ $(> x_*)$ climbs the potential, slowly passes the non-zero maximum position, and then converges to $x=0$. The graph of $x$ as a function of $t$ exhibits a two-step 
relaxation behavior with a long plateau. Furthermore, if $\varepsilon =0$, the particle converges to the position $x_*$
and cannot approach the origin. 
This qualitative change in the behavior of the system as a 
function of $\varepsilon$ occurring at $\varepsilon=0$ is analogous to that 
exhibited by glassy systems at the non-ergodic transition.

%
%

Clearly, this simple Newtonian system has no direct relation with glassy systems.
Nevertheless, it might be possible that there exists a variable representing 
the time correlation function for a glassy system that 
obeys an equation similar to that in the 
example discussed above. Such a variable would provide a characterization of 
the non-ergodic transition in that glassy system. 
Beginning from this speculation, in this paper, we determine such a variable 
and derive its evolution equation from a many-body Langevin model, employing several assumptions. We demonstrate that the obtained equation indeed does
exhibit a non-ergodic transition.

\section{Model}

%
%

We investigate a system consisting of $N$ identical Brownian particles 
suspended in a liquid. We denote the volume of the system by $V$ and the
temperature by $T$, and consider the thermodynamic limit, represented by $V \to \infty$ and $N \to \infty$, with the density $\bar \rho=N/V$ fixed.
Let $\bv{r}_i(t)$ be the position of the $i$-th particle,
with $i=1,2,\cdots N$. We assume that the $i$-th particle interacts with 
the $j$-th particle through some interaction potential 
$U(|\bv{r}_i-\bv{r}_j|)$.  The motion of each particle is described 
by a Langevin equation with a friction constant $\gamma$. Then, 
the evolution equation of the 
fine-grained density field, $\rho(\bv{r},t) \equiv \sum_{i=1}^N 
\delta(\bv{r}-\bv{r}_i(t))$, 
is given by
\begin{eqnarray}
\frac{\partial \rho(\bv{r},t)}{\partial t} = 
 \bv{\nabla} \cdot 
& & \left[ 
 \frac{1}{\gamma}\int d^3\bv{r'}\rho(\bv{r'},t)\rho(\bv{r},t) 
\bv{\nabla}U(|\bv{r}-\bv{r'}|) \right.
\nonumber  \\
& & + \left. \frac{T}{\gamma} \bv{\nabla} \rho(\bv{r},t) 
 +\sqrt{\frac{2T}{\gamma} \rho(\bv{r},t) } \bv{\Xi}(\bv{r},t) \right],
\label{Jevolve}
\end{eqnarray}
where $\bv{\Xi} (= (\Xi_x, \Xi_y, \Xi_z))$ is zero-mean Gaussian white noise that satisfies
\begin{eqnarray}
\langle \Xi_{\alpha}(\bv{r},t) \Xi_{\beta}(\bv{r'},s)\rangle 
= \delta_{\alpha \beta}\delta(t-s)\delta(\bv{r}-\bv{r'}).
\end{eqnarray}
This evolution equation can be derived exactly from the Langevin equation describing the motion of $N$ particles \cite{dean}. 
Explicitly, we choose the form of the interaction potential as
\begin{eqnarray}
\hat U(k)=
\frac{ E }{\lambda_1( k^2+\lambda_1^2)} 
- \frac{2k  E \sigma }{(k^2+\lambda_2^2)^2},
\label{eq:toypot}
\end{eqnarray}
where $k=|\bv{k}|$.
 Throughout this paper, $\hat f(\bv{k})$ is used to represent the Fourier transform of $f(\bv{r})$: $\hat f(\bv{k}) =
\int d^3 \bv{r}\e^{i\bv{k}\cdot\bv{r}}f(\bv{r})$. 
All quantities in this system are converted
into dimensionless forms by setting $\lambda_1=E=\gamma=1$. Further, 
for the specific 
example that we analyze numerically, we consider the case in which 
$\sigma=8.0$ and $\lambda_2=\lambda_1$. With this choice, $U(r)$ has a minimum
value $-1.18$ at $r=0.09$, while $U(r) \to \infty$ as $r \to 0$ and 
$U(r) \to 0$ as $r \to \infty$.
Explicitly one can obtain 
\begin{eqnarray}
U(r) = 
\frac{\e^{-r}}{4 \pi r}-\frac{4}{\pi^2}
\lr{1+\frac{r}{2}\frac{d}{dr}}\frac{\e^{-r}{\rm Ei}(r)-\e^r{\rm Ei}(-r)}{r},
\end{eqnarray}
where ${\rm Ei}(r)$ represents an integral exponential function defined as ${\rm Ei}(r) = \int_{-\infty}^r dt \e^t/t$.
We also fix the density $\bar \rho$  as 
$\bar \rho=4.436$.
The temperature $T$ is regarded as a control parameter. 

%
%

For later convenience, we
define the quantity $\Psi(\bv{k},\bv{k'}; t,s) 
\equiv {\hat \delta \rho}(\bv{k},t){\hat \delta \rho}(\bv{k'},s)/\bar \rho^2 $, where $\delta \rho(\bv{r},t) \equiv \rho(\bv{r},t)-\bar \rho$.
The statistical average of $\Psi(\bv{k},\bv{k'}; t,s)$ provides the two-point correlation
function $\hat C(k,t)$ 
according to the relation 
\begin{equation}
\bra \Psi(\bv{k},\bv{k'}; t,s) \ket
=(2\pi)^3 \hat C(k,t-s) \delta(\bv{k}+\bv{k'}),
\end{equation}
where we have assumed that the statistical properties of density fluctuations are symmetric with respect to spatial translations, rotations, and temporal translations.
Although the last assumption is not valid
below the glass transition temperature, we focus on the higher 
temperature regime, in which stationarity holds. Now, following 
the motivation of this study, we wish to obtain a 
differential equation that determines the time dependence of the 
correlation function $ \hat C(k,t)$. 
For this purpose, 
as a trial, let us first consider the quantity $\partial^2 \hat C(k,t)/\partial t^2$. 
This quantity can be expressed in terms of $\hat C(k,t)$ and a four-point correlation function. 
Then, writing this equality, we obtain an evolution equation for $\hat C(k,t)$ by making the simplest approximation of omitting 
the third- and fourth-order cumulant terms of this four-point 
correlation function.  
The equation obtained with this approximation represents a mean field 
theory in the sense that fluctuation effects of $\Psi$ are 
ignored when we consider its average value.
As far as we know, this naive approximation was not employed in previous 
studies. It might be important to find some connection with more traditional approximations such as a Kirkwood superposition approximation.
With this approximation, we obtain the following  evolution
equation for $\hat C(k,t)$:
\begin{eqnarray}
& &\frac{\partial^2 \hat C(k,t)}{\partial t^2}
=\lr{\frac{\bar \rho}{\gamma}\hat U(k)
+\frac{T}{\gamma}}^2 k^4\hat C(k,t)\nonumber \\
& & 
+\frac{\bar \rho^2}{\gamma^2}
\int \frac{d^3\bv{k}_1}{(2\pi)^3}
\hat C(k_1,t)\hat C(|\bv{k}-\bv{k}_1|,t)
\nonumber\\
&  &   
\left[\hat U(k_1)^2\lr{\bv{k}\cdot\bv{k}_1}^2
+\hat U(k_1)\hat U(|\bv{k}-\bv{k}_1|)
\lr{\bv{k}\cdot\bv{k}_1}
\lr{\bv{k}\cdot\lr{\bv{k}-\bv{k}_1}}\right],
\label{Ck}
\end{eqnarray}
with $t>0$. Here, we impose 
the boundary conditions  
$\lim_{t \to \infty}\hat C(k,t)=0$ and 
$\partial_t \hat C(k,t) |_{t=0+}= -k^2 T\bar \rho/\gamma$,
which can also be derived from  (\ref{Jevolve}). 

%
%

Here, it is expected that 
$\hat C(k,t)$ decays rapidly to zero in the limit $k \to \infty$
and exhibits no singularity as a function of $k$ for any $t$.
Therefore, we can expand $\hat C(k,t)$ in terms of the set of Hermite functions
$\phi_n(k)$, with $n=0,1,\cdots$  as 
$\hat C(k,t) = \sum_{n=0}^\infty C_{n}(t)\phi_n(k).$
Then, (\ref{Ck}) can be rewritten as
 \begin{eqnarray}
\sum_{m=0}^\infty G_{nm}\frac{d C_m(t)}{d t} &=& B_n(t), 
\label{eq:C} \\
\sum_{m=0}^\infty G_{nm}\frac{d B_m(t)}{d t} &=& \sum_{m=0}^\infty C_m(t) L_{nm}+ 
\sum_{m,l=0}^\infty M_{nml} C_m(t) C_l(t).
\label{eq:B}
\end{eqnarray}
Here, we have introduced the new variable $B_n (t)$ 
in order to obtain a first-order 
differential equation.
The coefficients $L_{nm}$ and $M_{nml}$ in (\ref{eq:B}) 
are determined by the system parameters. For
example, $L_{nm}$ is calculated as 
\begin{equation}
L_{nm} =
\frac{\bar \rho^2}{\gamma^2}\int_{-\infty}^{\infty} dk
\phi_n(k)\phi_m(k) \left(\hat U(k)+\frac{T}{\bar \rho}\right)^2,
\label{eq:Lnm}
\end{equation} 
while $G_{nm}$ in (\ref{eq:C}) and (\ref{eq:B}) is given by 
\begin{equation}
G_{nm} =
-\int_{-\infty}^{\infty} dk  \frac{1}{k^2+\Lambda^2} 
\phi_n(k)\phi_m(k),
\label{eq:Gnm}
\end{equation}
where we have introduced a cutoff parameter $\Lambda$
in order to allow expansion of 
the inverse of the 
Laplacian in terms of the Hermite functions. 
We take the limit $\Lambda \to 0$ after the transition temperature 
is calculated for a system with $\Lambda$ fixed. 
In the actual computations,
checking the temperature in the cases
$\Lambda^2=10^{-2}$, $10^{-3}$,  and $10^{-4}$,
we found that $\Lambda^2=10^{-3}$ is a sufficiently small value. 
The numerical values appearing in the analysis hereafter were obtained 
using $\Lambda^2=10^{-3}$.

%
%

Now we describe a further approximation.
Setting 
\begin{equation}
\bv{u}=(C_0,C_1,\cdots,C_{K-1}, B_0,\cdots,B_{K-1}),
\end{equation} 
we approximate  (\ref{eq:C}) and (\ref{eq:B}) by the following 
ordinary differential equation:
\begin{equation}
{\hat G}\der{\bv{u}}{t}
={\hat \Sigma}\bv{u} +{\hat \Theta}(\bv{u},\bv{u}).
\label{eq:u}
\end{equation}
Here, the $2K \times 2K $ matrices $\hat G$ and $\hat \Sigma$ are 
defined as $\hat G_{nm} =\hat G_{K+n,K+m}=G_{nm}$ and 
$\hat \Sigma_{n,K+m}=\delta_{nm}$, $\hat \Sigma_{K+n,m}=L_{nm}$, for $1 \le n,m \le K$
(The other components are zero).
Also, ${\hat \Theta}$ is 
determined from $M_{nml}$ and represents a map from two copies of 
$\bv{u}$ to a vector. We chose $K=100$ for the analysis reported
below. (We checked the $K$ dependence of $L_{nm}$, $G_{nm}$ and 
$M_{nml}$, and confirmed that $K=100$ is sufficiently large.)

\section{Analysis}

%
%

We investigate the system behavior as we change the temperature $T$. 
We first focus on the behavior of solution trajectories of 
(\ref{eq:u}) near the origin, $\bv{u}=0$. When $|\bv{u}| $ is 
sufficiently small, this behavior is approximated by the linear 
equation obtained by omitting the nonlinear term in  (\ref{eq:u}). 
Then, calculating the eigenvalues and eigenvectors of $\hat \Sigma$,
we can classify the solution trajectories. In the high temperature 
limit, from the form of $L_{nm}$, it is easily found that all 
trajectories either approach or move away from the origin exponentially, 
because  the eigenvalues consist of $K$ pairs of positive and negative 
numbers. We found in the numerical computations that when the temperature
is decreasing there exists a temperature $T_0$ below which one pair of 
eigenvalues becomes zero. (We have $T_0$=19.535 in the 
example we study.) Thus, the solution trajectories in the $2K$-dimensional 
phase space undergo qualitative change when the temperature
passes through $T_0$. The corresponding 
eigenvectors at $T_0$, which we denote $\bv{\Phi}_{00}$
and $\bv{\Phi}_{01}$, play a prominent role in the subsequent analysis.

%
%

It is known that this type of qualitative change near $T=T_0$ can be 
described by a nonlinear differential equation for the amplitudes of 
these zero eigenvectors \cite{kuramoto}. We now attempt to derive such 
an equation. Let ${\hat \Sigma_0}$ be ${\hat \Sigma}$ at $T=T_0$. Then, 
we find that $\bv{\Phi}_{00}$ and $\bv{\Phi}_{01}$ satisfy the relations
\begin{eqnarray}
\hat \Sigma_0 \bv{\Phi}_{00} &=& 0, \\
\hat \Sigma_0 \bv{\Phi}_{01} &=& \bv{\Phi}_{00}.
\end{eqnarray}
Now, setting $\epsilon =(T-T_0)/T_0$, we consider the solution 
trajectories satisfying $\bf{u} \to 0$ in the limit $t \to \infty$ for 
the system with small positive $\epsilon$. For such trajectories, we 
expand $\bv{u}$ in eigenvectors of $\hat \Sigma_0$ as
\begin{equation}
\bv{u}(t) = A_1(t) \epsilon \bv{\Phi}_{00}
+A_2(t) \epsilon^{3/2} 
\bv{\Phi}_{01}+ \bv{\eta}(A_1(t),A_2(t)),
\label{A:def}
\end{equation}
where $\bv{\eta}$ represents the contribution to $\bv{u}$ that 
is not from the zero eigenvectors, and the amplitudes of the other 
eigenvectors quickly decay to the values determined by $A_1$ and $A_2$. 
Then, the time evolution of the variables $A_1$ and $A_2$ is described 
by autonomous equations of the form 
\begin{equation}
\frac{d A_i(t)}{d t}=\Omega_i(A_1(t),A_2(t)).
\label{A:eq}
\end{equation}

%
%

Now, substituting (\ref{A:def}) into (\ref{eq:u}) and using expansions 
\begin{eqnarray}
\bv{\eta} & = & \epsilon^{3/2}\bv{\eta}^{(3/2)}
+\epsilon^{2} \bv{\eta}^{(2)}
+\epsilon^{5/2} \bv{\eta}^{(5/2)}+ \cdots,
\label{rho:exp} \\
\Omega_i & = & \epsilon^{1/2} \Omega_{i}^{(1/2)}
+\epsilon^{1}\Omega_{i}^{(1)}+\epsilon^{3/2} \Omega_{i}^{(3/2)}+\cdots,
\label{A:exp}
\end{eqnarray}
we can determine the functional forms of $\bv{\eta}(A_1,A_2)$ and 
$\Omega_i(A_1,A_2)$ using a perturbative method. Note that 
the dependences on $\epsilon $ in (\ref{A:def}), (\ref{rho:exp}), and 
(\ref{A:exp}) are chosen so that the perturbative expansion can be 
carried out in a systematic manner. For example, all the terms 
proportional to $\epsilon^{3/2}$ in the expression obtained with the 
above procedure  yield 
\begin{equation}
{\hat G}  \bv{\Phi}_{00}\Omega_1^{(1/2)}(A_1,A_2)
=A_2 \bv{\Phi}_{00} + {\hat \Sigma_0}\bv{\eta}^{(3/2)}(A_1,A_2).
\label{1st}
\end{equation}
Because $\hat \Sigma_0$ has a zero eigenvalue, it is not invertible. 
Thus, when we regard  (\ref{1st}) as a linear equation for 
$\bv{\eta}^{({3/2})}$, there are either no solutions or an infinite 
number of solutions. In order to carry out the perturbative calculation
systematically, we impose the solvability condition under which 
$\bv{\eta}^{(3/2)}$ can be obtained uniquely. 
This solvability condition 
determines $\Omega_1^{(1/2)}$. Applying similar procedures to each 
of the equations appearing successively, which are proportional to 
$\epsilon^{2}$, $\epsilon^{5/2}$, $\cdots$, we derive $\Omega_1^{(1)}$, 
$\Omega_2^{(1/2)}$, $\Omega_1^{(3/2)}$, $\Omega_2^{(1)}$, $\cdots$
in an iterative manner. Combining the contributions up to 
$\Omega_{1}^{(2)}$ and $\Omega_{2}^{(3/2)}$, we obtain
\begin{eqnarray}
\frac{d A_1}{d t} &=& a A_2,  
\label{A1:final}   \\
\frac{d A_2}{d t} &=& b_1 A_1+b_2 A_1^2+b_3 A_1^3+b_4 A_2^2, 
\label{A2:final}
\end{eqnarray} 
where each coefficient can be calculated as a function of $T$. For the 
example we analyzed numerically, we found
$a=-0.061186 \epsilon^{1/2}$, 
$b_1=-1.108178\epsilon^{3/2}$,
$b_2=-0.000566\epsilon^{1/2}$,
$b_3=-0.011904 \epsilon^{3/2}$ and  $b_4=0.836102 \epsilon^{3/2}$. 
(The details of the calculation will be reported in another paper 
\cite{next}.)

%
%

\begin{figure}[htbp]
\begin{tabular}{ccc}
\begin{minipage}{0.33\hsize}
\begin{center}
\includegraphics[width=\hsize]{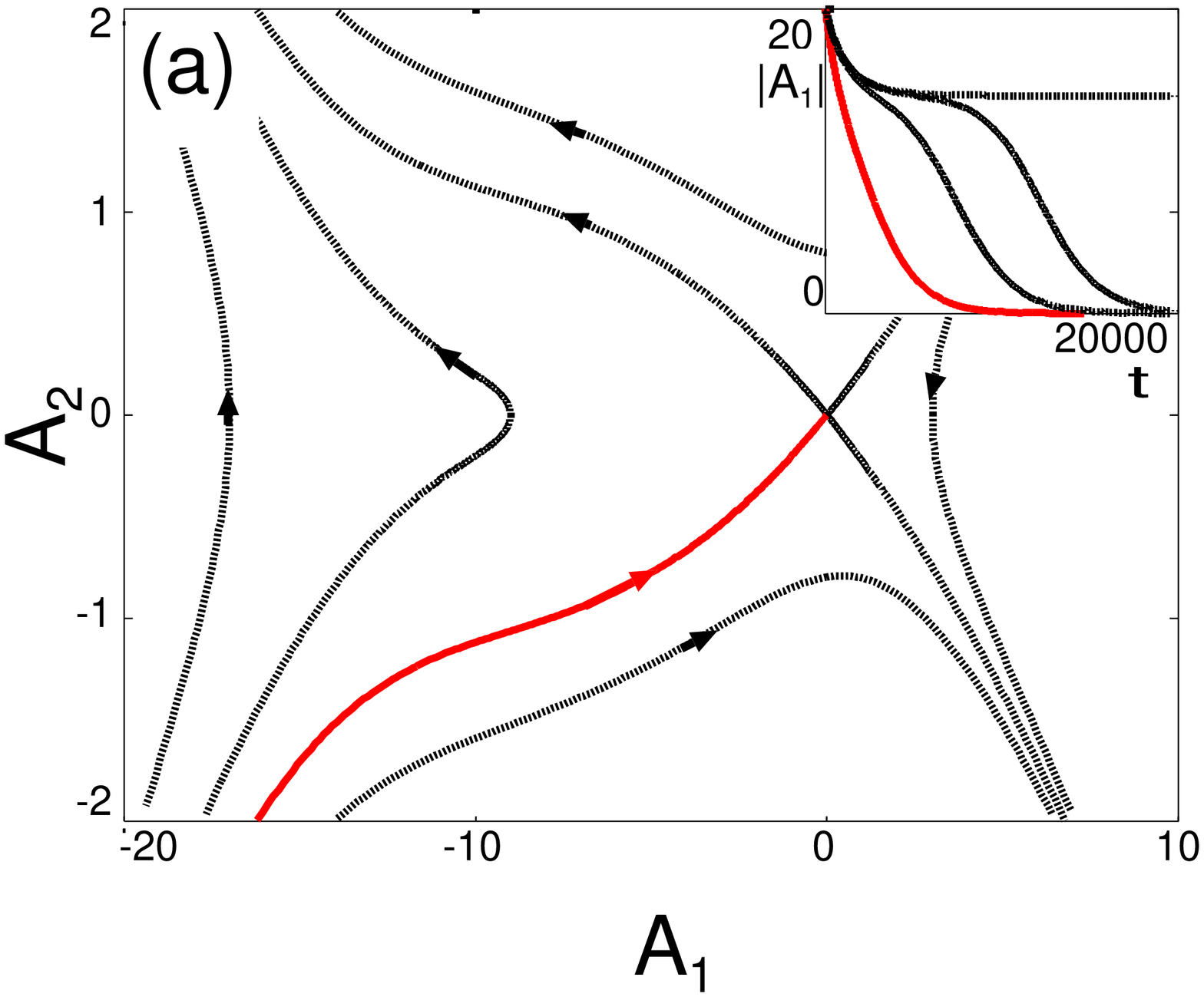}
\end{center}
\end{minipage}
\label{phase}
\begin{minipage}{0.33\hsize}
\begin{center}
\includegraphics[width=\hsize]{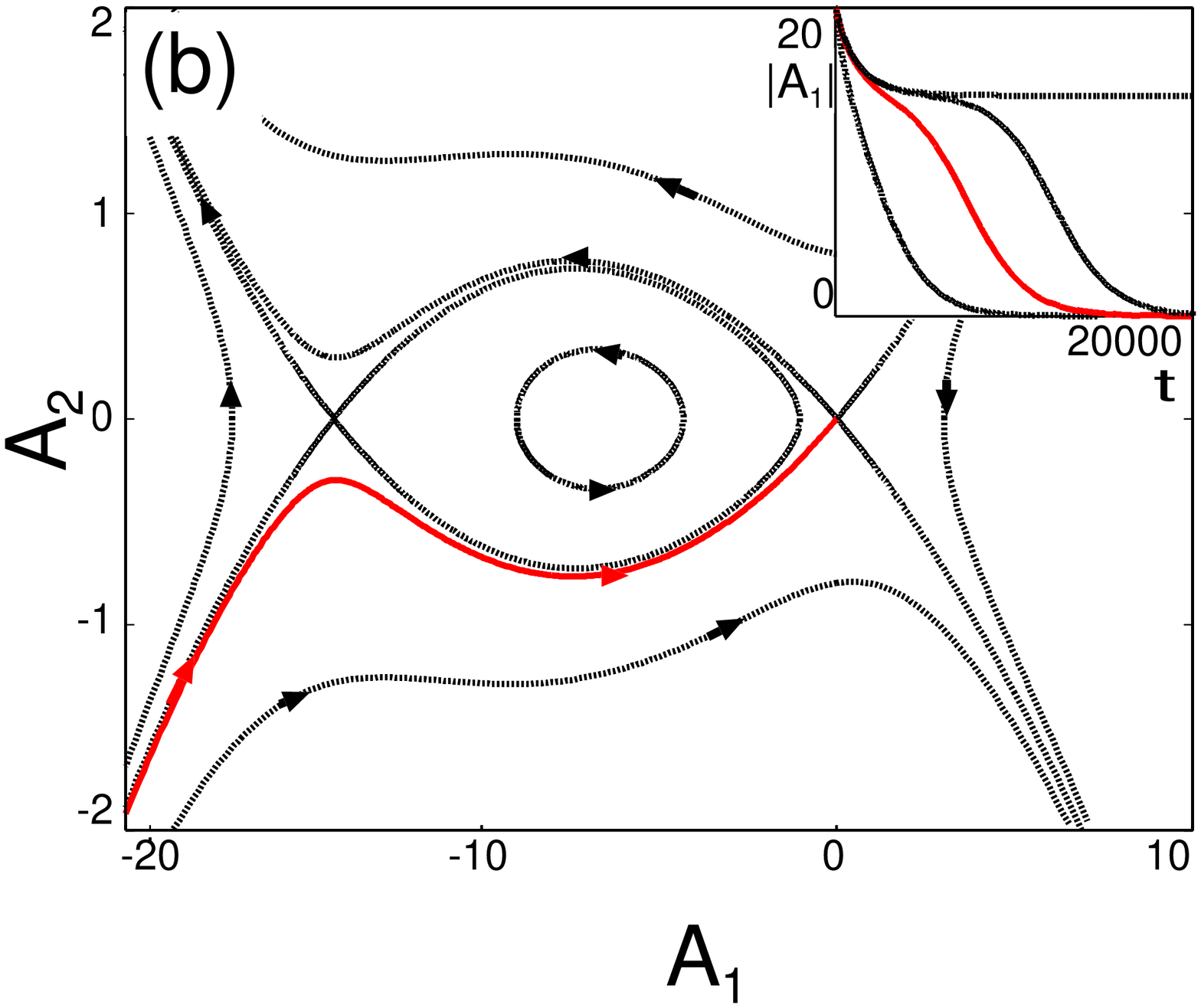}
\end{center}
\end{minipage}
\begin{minipage}{0.33\hsize}
\begin{center}
\includegraphics[width=\hsize]{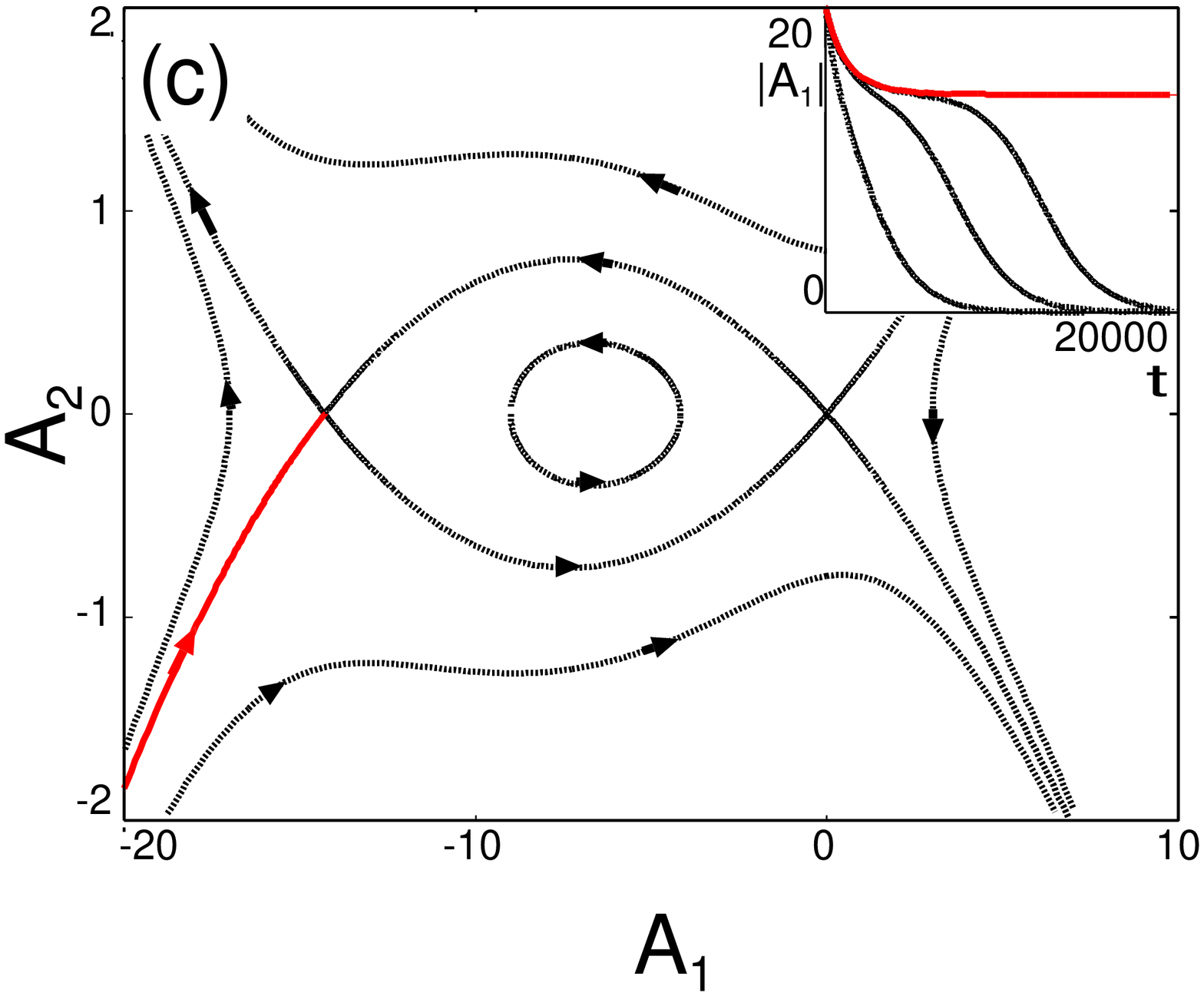}
\end{center}
\end{minipage}
\end{tabular}
\caption{ Solution trajectories of (\ref{A1:final}) and (\ref{A2:final}) in $(A_1,A_2)$ space for three values of $\epsilon$. 
(a) $\epsilon=2.520 \times 10^{-3}$, (b) $\epsilon=2.294 \times 10^{-3}$,
and (c) $\epsilon=2.285 \times 10^{-3}$. 
The red curves represent physical solutions that satisfy $\lim_{t\to\infty}A_i =0$.
Inset: $A_1$ as a function of $t$. 
The red curves in the insets  correspond to the red trajectories.
}
\end{figure}

Using this result, we plot trajectories of $(A_1(t), A_2(t))$ in 
Fig. 1 
for three values of $\epsilon$. It is seen that for
$\epsilon=2.520 \times 10^{-3}$, the trajectory approaches the origin 
with no particularly notable features. By contrast, for 
$\epsilon=2.294 \times 10^{-3}$, there appear two fixed points, and the 
trajectory approaches one of these. However, in this case, 
it eventually begins to move away from this point and converges to the 
origin in the $t \to \infty$ limit. 
As shown in the inset of Fig. 1,
 the part of the trajectory 
near the fixed point corresponds to the plateau in the graph of $A_1$ 
as a function of time. Finally, for $\epsilon=2.285 \times 10^{-4}$, 
we find that the trajectory converges to the non-zero fixed point as $t \to \infty$. 
The appearance of the connection between this new fixed point and the origin is called a {\it saddle connection bifurcation} in dynamical system theory. The inset
of Fig. 1 shows that 
the saddle connetion bifurcation in the two-dimensional system studied here 
corresponds to the  non-ergodic 
transition of the time correlation function. 

\section{Concluding remark}

%
%

We have elucidated the nature of a non-ergodic transition using a 
method of dynamical system reduction to treat the time correlation function.
The key idea employed in our analysis 
is to focus on the zero eigenvectors ${\bv \Phi}_{00}$
and ${\bv \Phi}_{01}$ at the temperature $T_0$. Using these 
eigenvectors, we can obtain a simplified description of 
the large-dimensional dynamical system (\ref{eq:u}) in the form of the two-dimensional system represented by (\ref{A1:final}) and (\ref{A2:final}) near the temperature $T_0$. 
We find that the non-ergodic transition in the original system is manifested 
as the saddle connection bifurcation in the two-dimensional system.
Before ending this paper, we make three important remarks.

%
%

First, 
our theoretical result for the system with the potential (\ref{eq:toypot}) 
has not been compared with numerical experiments yet.
The comparison will be a future work.
Here, we note that our method can be applied to Langevin equations with other interaction potentials.
As an example, for a model recently studied in Ref. 
\cite{dotsenko}, we confirmed the non-ergodic transition by 
both our method and a direct numerical experiment, though there is no 
precise quantitative correspondence between them.
We also note that we do not understand the dependency of the results on potential forms. 


The second remark is that 
(\ref{Jevolve}) takes the same form 
as the density evolution equation in the over-damped limit obtained with
the dynamical density functional method, where the potential function in 
(\ref{Jevolve}) is replaced with the direct correlation function \cite{das}.
It is thus expected that our method 
can also be applied to this type of density evolution, though the temperature dependence becomes complicated.

%
%


%
%

Finally, recall that the equation we analyzed, (\ref{eq:u}),
was obtained by ignoring fluctuations of $\Psi(\bv{k}, \bv{k'};t,s)$. 
Thus, it is important to extend our treatment to one that accounts for 
fluctuations, in which  (\ref{eq:u}) is derived 
as the zero-th order description.
This problem is related to that of the growth of the length scale
that appears near the glass transition point, which has been studied recently \cite{chi4, BB}. 
We are now in the process of developing such a theory.

%
%

\ack
The authors thank 
S.-H Chong, 
K. Hukushima, 
K. Kawasaki, 
W. Kob, 
K. Miyazaki, and 
A. Yoshimori 
for useful comments 
on this study.
This work was supported by a grant from the Ministry of Education, 
Science, Sports and Culture of Japan (No. 16540337).

\end{document}